\documentstyle[12pt]{article}

\def\lsim{\mathrel{\mathpalette\vereq<}}
\def\gsim{\mathrel{\mathpalette\vereq>}}
\makeatletter
\def\vereq#1#2{\lower3pt\vbox{\baselineskip1.5pt \lineskip1.5pt
\ialign{$\m@th#1\hfill##\hfil$\crcr#2\crcr\sim\crcr}}}
\makeatother

\begin{document}

\begin{titlepage}
\begin{center}
\today     \hfill    LBNL-39865 \\
~{} \hfill UCB-PTH-96/55  \\
~{} \hfill IASSNS-HEP-96/106 \\
~{} \hfill hep-ph/9701286 \\

\vskip .25in

{\large \bf Building Models of Gauge Mediated Supersymmetry Breaking
\\ Without A Messenger Sector}\footnote{This work was
supported in part by the Director, Office of
Energy Research, Office of High Energy and Nuclear Physics, Division of
High Energy Physics of the U.S. Department of Energy under Contracts
DE-AC03-76SF00098 and DE-FG-0290ER40542
and in part by the National Science Foundation under
grant PHY-95-14797.  NAH was also supported by NSERC,
JMR by the W. M. Keck Foundation, and HM by the Alfred P. Sloan
Foundation.}

\vskip 0.3in

Nima Arkani-Hamed,$^{1,2}$ John March-Russell,$^{3}$
and Hitoshi Murayama$^{1,2}$

\vskip 0.1in

{{}$^1$ \em Theoretical Physics Group\\
     Earnest Orlando Lawrence Berkeley National Laboratory\\
     University of California, Berkeley, California 94720}

\vskip 0.1in

{{}$^2$ \em Department of Physics\\
     University of California, Berkeley, California 94720}

\vskip 0.1in

{{}$^{3}$ \em Institute for Advanced Study\\
    School of Natural Sciences\\
    Princeton, NJ 08540}

\end{center}

\vskip .1in

\begin{abstract}
We propose a general scheme for constructing models in which the
Standard Model (SM) gauge interactions are the mediators of
supersymmetry breaking to the fields in the supersymmetric SM,
but where the SM gauge groups couple directly to the sector which
breaks supersymmetry dynamically.  Despite the direct coupling, the
models preserve perturbative unification of the SM gauge coupling
constants.  Furthermore, the supergravity contributions to the
squark and slepton masses can be naturally small, typically being
much less than 1\% of the gauge mediated (GM) contributions.  Both
of these goals can be achieved without need of a fine-tuning or a
very small coupling constant.  This scheme requires run-away
directions at the renormalizable level which are only lifted by
non-renormalizable terms in the superpotential.  To study the
proposed scheme in practice, we develop a modified class of models
based on $SU(N)\times SU(N-1)$ which allows us to gauge a $SU(N-2)$
global symmetry.  However, we point out a new problem which can
exist in models where the dynamical supersymmetry breaking sector and
the ordinary sector are directly coupled -- the
two-loop renormalization group has contributions which can induce
negative $({\rm mass})^2$ for the squarks and sleptons.  We
clarify the origin of the problem and argue that it is likely to
be surmountable.  We give a recipe for a successful model.
\end{abstract}

\end{titlepage}

\newpage

\section{Introduction}
\setcounter{footnote}{0}
\setcounter{equation}{0}

Low-energy supersymmetry is widely regarded as the most promising
stabilization mechanism for the hierarchy between the Planck scale
and the weak scale (for a review see
\cite{INS}).  However, supersymmetry by itself does not necessarily
generate a hierarchy.  If the hierarchy is to be explained, it most
likely has to result from the breaking of supersymmetry by
non-perturbative dynamics so that the existence of a small scale is
naturally understood by dimensional transmutation  \cite{Witten}.
This motivates the study of dynamical supersymmetry breaking in
supersymmetric gauge theories \cite{ADS}.

The scheme most commonly considered in the literature
assumes that supersymmetry is broken dynamically in a hidden sector
whose coupling to the Standard Model (SM) is solely due to Planck-scale
suppressed interactions \cite{hidden}.  The motivation for such ideas
is increased since gravity automatically provides such a
coupling, and, furthermore it seems that string theory naturally
accommodates hidden sectors.  However, there has been
a growing realization that string theory, or supergravity alone, does
not automatically possess a phenomenologically viable mechanism
to maintain the degeneracy between sfermions of the same
gauge quantum numbers.  This degeneracy is the simplest way in which
the stringent constraints from flavor-changing neutral currents and
other rare processes can be met \cite{DG}.

However, if supersymmetry breaking is dominantly communicated to
the supersymmetric standard model (SSM) by the SM gauge
interactions themselves, then the required degeneracy
is automatic, given that the ever-present supergravity contributions
are negligible.  This highly appealing mechanism of gauge mediation
(GM) guarantees enough
suppression of flavor-changing neutral currents \cite{oldpapers}.

Explicit, and quite complete, models based on gauge mediation have
been constructed in the past few years \cite{DN,DNS,DNNS}.  These
models achieve the gauge mediation of supersymmetry breaking in the
following way.  The sector which breaks supersymmetry dynamically (the
DSB sector) has a non-anomalous global symmetry which is weakly gauged
(the messenger gauge group).  The exchange of the messenger gauge multiplet
induces supersymmetry breaking effects in a second sector (the
messenger sector).  Concretely, this messenger sector contains at
minimum a gauge singlet chiral superfield $S$, and a set of fields
which transform under the standard model gauge groups in a vector-like
manner (messenger fields).  The supersymmetry breaking effects induce
expectation values for both the $A$- and $F$-components of $S$.  The
coupling of $S$ to the vector-like messenger fields then in turn
induces, via these messenger fields, the SSM gaugino and soft scalar
masses via one- and two-loop diagrams, respectively.  We will refer to
this class of models as Original Gauge Mediation (OGM) Models.

Even though the final stage of SM gauge mediation is itself quite
appealing, the OGM models seem rather contrived in their use of a
cascade of different interactions to communicate supersymmetry
breaking to the SSM.  In fact, the authors of Refs.~\cite{DN,DNS,DNNS}
were forced to introduce the messenger sector to ``insulate''
the standard model from the dynamical
sector so as to maintain the perturbative SM gauge coupling
unification as indicated by the data from the LEP/SLC experiments.  If
one couples the DSB to the SSM sector directly via the standard model
gauge interactions, there tends to be a large multiplicity of new fields
charged under the SM, so that the SM gauge coupling
constants reach their Landau poles well before unification.
Moreover, if one wishes to avoid a color-breaking minimum of the
potential \cite{ACHM,DDR}, the messenger sector has to be even more
complicated than originally thought.  Also as we will discuss
in Section~2, the OGM models suffer from a number of phenomenologically
uncomfortable features.  Thus, given the recent
progress in understanding supersymmetric gauge theories (for a review
see \cite{IS}), it is natural to ask whether one can find a more
elegant scheme based on the new DSB models recently constructed
\cite{DNNS,HM,PT1,PS,IY,IT1,PST1,CRS,PST2,CRSL,IT2}.

In this paper, we study a general scheme in which the DSB
sector is coupled directly to the SM gauge interactions,
while maintaining all the attractive features of the basic GM
idea.  Beyond the simplicity of the direct coupling, we aim to
achieve three other ambitious goals: (1) the preservation of perturbative
gauge unification, (2) the natural suppression of supergravity contributions
to the sfermion masses so as to maintain degeneracy, (3) no fine-tuning
of parameters or very small coupling constants.  The basic idea is to
find a model which has a classical flat direction, $X$, not lifted at
the renormalizable level, but which allows a non-renormalizable operator
in the superpotential that does lift $X$.  The direction $X$ then acquires
a large expectation value while the theory maintains only a small
vacuum energy.  This results in a natural hierarchy
$\langle F_{X} \rangle \ll \langle X \rangle^{2}$.  If it is arranged
that the fields charged under the SM acquire masses due to
$\langle X \rangle$, their contribution to the running of the gauge
coupling constants appears only above $\langle X\rangle$ and hence
perturbative gauge unification can be preserved.  On the other hand, the
superparticle masses generated by GM are proportional
to the ratio $\langle F_{X} \rangle / \langle X \rangle$ and can be kept
at the phenomenologically desired value $\langle F_{X} \rangle /
\langle X \rangle \sim 10^4$~GeV.
As an added bonus, we point out that such a scheme overcomes many of
the phenomenologically undesirable characteristics of the OGM models.
A discussion of these features is contained in Section~6.


In fact, Poppitz and Trivedi \cite{PT2} have recently built a
model along the general lines stated above in which the
dynamical sector and the standard model groups are directly coupled,
while keeping the gauge coupling constants perturbative up to the
GUT-scale.  Unfortunately their model suffers from two severe problems.
The first is that supergravity mediated soft masses are of the same order
of magnitude as the gauge-mediated soft terms, and thus sfermion
degeneracy is not naturally guaranteed, spoiling the original
motivation for GM.

We demonstrate that this problem can be overcome by developing a modified
class of DSB models based on a $SU(N)\times SU(N-1)$ gauge group, that
allows a $SU(N-2)$ global symmetry to be weakly gauged.  To achieve this
it is necessary to add new fields to the models of Poppitz, Shadmi and
Trivedi \cite{PST1}.  It is non-trivial that at the quantum level
the resulting additional classical flat-directions are lifted, and DSB is
maintained.  We perform detailed analyses of the potential and the
particle spectrum, and it is demonstrated that this class of models
achieves the three goals listed above.  In particular, even though
$\langle X \rangle$ is close to the GUT-scale, the modified
models which we present naturally suppress the supergravity
contributions to the sfermion masses.

However, as a consequence of our analysis, we discover a new problem
that unfortunately afflicts both our models and those of Poppitz and
Trivedi.  Explicitly, there exist at two-loops, contributions to the
renormalization groups equations of SSM scalar $({\rm mass})^2$ that
drive them negative at the weak scale.  This problem appears to be
model-dependent, and hinges upon the details of exact numerical
coefficients.  It occurs only in models
where there are light ($\sim 10^4$~GeV) fields charged under the standard
model gauge groups which directly originate in the DSB sector, and which
thereby acquire a large soft-SUSY-breaking scalar mass of order
$(10^4~\mbox{GeV})^2$.  We believe that this is not a generic
consequence of our proposed scheme, and given the
rapidly growing list of DSB models,  theories
fulfilling the three goals without
generating particles charged under the SM gauge groups with large
SUSY-breaking masses will be found.

The organization of the paper is as follows.  In Section~2 we review
the basics of the original gauge mediation models, and discuss a number
of their phenomenologically less desirable features, some of which do
not seem to be widely appreciated.  In Section~3 we outline
previous attempts at simplifying the structure of the OGM models.
Section~4 explains in greater detail the structure of our
proposed scheme, readers who are only interested in the
requirements on DSB models which allow the direct coupling to the ordinary
sector should go directly to this section.
Section~5 contains the $SU(7)\times SU(6)$ model
and its analysis, and demonstrates that it fulfills our three goals.
In Section~6 we address the phenomenologically favorable aspects
of this general scheme.  Section~7 discusses the problem that our
$SU(N)\times SU(N-1)$ models, as well as those of Poppitz and Trivedi,
have with large negative contributions to the SSM sfermion $({\rm mass})^2$.
We conclude in Section~8, while a long appendix contains many details of the
modified $SU(N)\times SU(N-1)$ models.

\section{Original Gauge Mediation Models}
\label{sec:original}
\setcounter{footnote}{0}
\setcounter{equation}{0}

In this section we review the OGM models of Dine, Nelson and
collaborators, and then go on to discuss various phenomenological
problems of these models.
Let us first review the mechanism of gauge
mediation itself.  Consider $N$ vector-like multiplets $q_i$ and
$\bar{q}_i$ all transforming as ${\bf 3}+{\bf 3}^{*}$ under color
$SU(3)$.  Suppose that they acquire an invariant supersymmetric
mass
\begin{equation}
	W = \langle S\rangle \bar{q}_i q_i
\end{equation}
due to the $A$-component of a chiral superfield $S$ gaining
an expectation value, as well as a supersymmetry-breaking bilinear mass
term
\begin{equation}
	V = \langle F_S\rangle (\tilde{\bar{q}}_i\tilde{q}_i + \mbox{c.c.})
\end{equation}
in the potential, due the $F$-component of $S$ also gaining an
expectation value as a result of some dynamics.  Here
$\tilde{\bar{q}}$ and $\tilde{q}$ represent the scalar
components of the corresponding chiral superfields.  Integrating out
these vector-like multiplets generates gluino and squark masses,
\begin{eqnarray}
	M_{\tilde{g}} & = & \frac{N\alpha_{s}}{4\pi} B,
	  \\
	m^{2}_{\tilde{q}} & = & 2 N C_{2}
		\left(\frac{\alpha_{s}}{4\pi}\right)^{2} B^{2} ,
\end{eqnarray}
where $B \equiv \langle F_S\rangle/ \langle S\rangle$.
Here $C_{2} =4/3$ is the second-order Casimir invariant for the relevant
representation.  Similar results hold for the SM states charged under
$SU(2)$ and $U(1)$ once vector-like messengers carrying these quantum
numbers are similarly included.

There are two important points to note about this mechanism.  First, for
$N\simeq 1$, the gaugino and scalar masses are generated with comparable
magnitudes, which is phenomenologically desirable.  Second, the scalar
masses are flavor-blind, {\it i.e.}\/ the same for scalars of the same
gauge quantum numbers in different generations.  This is vital if one
wishes to suppress the phenomenologically dangerous flavor-changing
effects mediated by sfermion loops.  Finally, the typical size of the
supersymmetry breaking for the messengers is given by $\langle F_S\rangle$
in such models.  It is noteworthy that this size of supersymmetry
breaking is itself not important in estimating the squark, slepton or gaugino
masses; rather, they are functions of the ratio $B$ only.  For
phenomenologically desirable gaugino and sfermion masses we therefore
require $B \simeq 60~{\mbox TeV}$, for $N=1$, with $B$
correspondingly reduced for larger values of $N$.

On the other hand, the gravitino mass is sensitive to the fundamental
size of supersymmetry breaking.  In the simplest and most successful OGM
models \cite{DNNS}, the DSB sector has a $U(1)$ global symmetry which
can be weakly gauged (messenger $U(1)$).  If extra fields are introduced
which are charged under this $U(1)$ and which also couple in the
superpotential to $S$, then it is possible at two-loops for the scalar
components of these charged fields to pick up soft $({\rm mass})^2$
terms, which in turn lead to the $A$ and $F$ components of $S$
gaining vacuum expectation values.  Thus, even if we assume
large values for the messenger $U(1)$ gauge coupling, the fundamental
scale of supersymmetry breaking must be $\Lambda\simeq 5,000$~TeV, or
larger, in these models.
Thus the gravitino mass, which is given by $m_{3/2} \simeq \Lambda^2/M_{*}$,
where $M_{*} = M_{\rm Planck}/\sqrt{8\pi}$ is the reduced Planck
mass, is not simply a function of $B$.

Before going on to attempts to simplify this structure,
we wish to review some of the physics problems of the OGM
models, not just related to aesthetics. \\[0.2cm]
{\bf $\mu$-problem.}
The so-called $\mu$-term, the invariant mass term for the
Higgs doublets in the superpotential, has to be of the same order as
the supersymmetry breaking parameters if we are to naturally explain
the stability of the weak scale, even though it is allowed by
supersymmetry.  Now, dynamical supersymmetry breaking can potentially
explain why $\mu$ is of order the weak scale if the $\mu$-term is also
generated dynamically.  However, the particular implementations of
this idea discussed so far look even more contrived than the OGM
models \cite{DNNS,DGP}.  This is due to a combination of different
problems.  The most naive attempts tend to generate too large a
supersymmetry breaking $\mu$-term (sometimes called $m_{3}^{2}$ or
$B\mu$) for an appropriate size of $\mu$.  A coupling to a singlet
Higgs field appears to be the next thing to try; however it usually
results in a light axion-like pseudo-Nambu-Goldstone boson in the
Higgs spectrum and is excluded by the $Z$-decay experiments \cite{DN}.
The reason for the existence of this light pseudo-scalar is an approximate
$R$-symmetry in the Higgs sector superpotential which exists for any
strictly trilinear couplings.  The $R$-symmetry is explicitly broken
by trilinear supersymmetry breaking terms, which are induced by
gaugino masses at one-loop; however their magnitudes turn to be too
small because of the limited amount of running between the messenger
scale and the weak scale in models with messengers at the
$10^4-10^5$~GeV scale.\\[0.2cm]
{\bf Electroweak symmetry breaking.}
It is an intersting fact that radiative electroweak symmetry breaking
can work within the OGM models.  Even though the logarithm
of the messenger scale to the weak scale is not large, the Higgs boson
mass squared $m_{2}^{2}$ is driven negative because the squarks acquire
much larger masses than the Higgs bosons in the gauge mediated scenarios.
However, the resulting mass is too negative
\cite{ACHM}, of the order of $m_{2}^{2} \simeq -(500$~GeV$^{2})$.
This is to be compared to the value required by
electroweak symmetry breaking: $\mu^{2} + m_{2}^{2} = - m_{Z}^{2}/2
= - (70$~GeV$^{2})$ for a moderately large $\tan\beta$.  It is possible
to fine-tune the negative mass-squared of the Higgs with a large positive
$\mu^{2}$, but a displeasing fine-tuning at the level of a percent
seems unavoidable. (And clearly this question is then linked with the
first question of how to generated the $\mu$-term.)\\[0.2cm]
{\bf Global minima of the messenger potential.}
The superpotential of the messenger sector in OGM models requires us
to sit at a local, rather than a global, minimum of the potential
\cite{ACHM,DDR}.  Specifically,
the vector-like messenger fields have a $D$-flat direction under
the standard model gauge group, which has a much lower energy than
the desired minimum.  One can
complicate the messenger sector by introducing another singlet or
other fields so as to make the desired minimum the global one, but this
certainly makes the model even more baroque.\\[0.2cm]
{\bf $ll\gamma\gamma$ events.}
It has been claimed that within the OGM-like models the lightest neutralino
or sleptons can decay into the gravitino and a photon/lepton (possibly
involving a cascade of decays) within the Fermilab collider detectors,
leaving a distinct experimental signature (of photon(s) plus missing
$E_T$) as compared to hidden sector supergravity mediated
models.  In particular the presence of two $ll\gamma\gamma$ events
within Run I of the Tevatron data has generated understandable interest.
Unfortunately, we are not aware of any consistent models
which actually give a low enough gravitino decay constant so as to
allow the lightest neutralino or slepton to decay inside the detector
\cite{Andre}.
The only candidate is a vector-like model directly coupled to the
messengers \cite{IT1}, which relies on a local minimum and a dynamical
assumption.\\[0.2cm]
{\bf Exotic stable particles.}
The DSB sector and messenger sectors tend to possess stable particles, some
of which are charged.  Even though it is just possible that they might
be dark matter candidates, they tend to overclose the Universe as
their mass scale is quite high \cite{DimoGP}.  Moreover, there are
very strong constraints on the abundance of stable charged particles
\cite{Esmailzadeh}.\\[0.2cm]
{\bf $R$-axion.}
Most of models which break supersymmetry dynamically have an exact but
spontaneously broken $U(1)_R$ symmetry.  They therefore produce a
massless Nambu--Goldston boson.  A combination of light meson decays,
beam dump experiments, quarkonium decay, the population of red
giants or white dwarfs in globular clusters, and the duration of
the supernova 1987A neutrino burst put lower bounds on the scale of
$U(1)_R$ symmetry breaking of around $10^{8}$~GeV.  However,
it was argued that the cosmological constant must be cancelled by
introducing a constant term in the superpotential which thereby breaks
the $U(1)_R$ symmetry explicitly.  This explicit breaking can generate
an $R$-axion mass which avoids astrophysical constraints \cite{BPR}.
It is, however, somewhat displeasing that the solution depends on
this mechanism of cancelling the cosmological constant, given that
we understand so little about why it is so small.   Other solutions to
the cosmological constant problem, such as no-scale supergravity, may
not solve the $R$-axion problem \cite{DNNS}.\\[0.2cm]
{\bf Cosmology.}
Even though it is correct that the OGM models do not suffer from the
Polonyi-problem \cite{BKN}, one of the major cosmological problems of the
supergravity-mediated hidden sector scenarios, there exist other
cosmological difficulties in these models.  Specifically, a gravitino with
mass of order 100~keV is expected in the OGM models.  This is in the
most disfavored mass range from the cosmological point of
view \cite{MMY}.  If the scenario is further regarded as a low-energy
limit of superstring theory, the string dilaton/moduli also have very
light masses of order 100~keV.  For such masses they are stable for
cosmological time scales and their coherent oscillations vastly
overclose the Universe by 15 orders of magnitude \cite{Andre}, unless a
period of keV-scale inflation is invoked with its attendent
problems.

\section{Previous Attempts}
\label{sec:previous}
\setcounter{footnote}{0}
\setcounter{equation}{0}

In this section we review some previous attempts to simplify the OGM
models discussed in Section~\ref{sec:original}.  The most interesting
of these in our opinion is due to Poppitz and Trivedi~\cite{PT2},
which we discuss in the next section.

Only a few attempts to couple the standard model gauge group
directly to the DSB sector have been made because
of the following problems.  In order for the dynamical sector to have
a large enough global symmetry, a subgroup of which is identified as
the SM gauge group, the gauge group in the dynamical sector tends to
become very large.  This results in a large number of extra matter
fields transforming non-trivially under the SM, leading to a Landau
pole for the SM gauge couplings only slightly
above the messenger scale.  Also the supergravity contribution to
the squark and  slepton masses must be small enough compared to the
contribution from GM to guarantee sufficient sfermion degeneracy -- the
main motivation for GM of supersymmetry breaking.

The existence of a class of models in which supersymmetry is
dynamically broken and which accomodate relatively large global
symmetries has long been known.  The $SU(2k+1)$ models with
an anti-symmetric tensor $A^{\alpha\beta}$ and $(2k-3)$
anti-fundamentals $\bar{F}_{\alpha}^{i}$, are the classic examples
constructed by Affleck, Dine, and Seiberg \cite{ADS} which break supersymmetry
dynamically.\footnote{This class of models is {\it non-calculable}\/
because they do not have classical flat directions along which one can
analyze the theory semi-classically.  However, the Witten index of
these models can be shown to vanish by adding a massive vector-like
field and so the theories are likely to break supersymmetry \cite{HM}.
Various subgroups of the models also allow models of supersymmetry
breaking, {\it e.g.}\/, $SU(2k)\times U(1)$,
$SU(2k-4)\times SU(5)\times U(1)$, or
$SU(2k-3)\times SU(4)\times U(1)$ \cite{CRS,CRSL}.}
The superpotential which lifts all classical flat directions, $W =
\lambda_{ij} A^{\alpha\beta} \bar{F}_{\alpha}^{i} \bar{F}_{\beta}^{j}$
can preserve an $SP(k-2)$ symmetry if the $\lambda_{ij}$ coupling
constant matrix is proportional to the symplectic matrix $J$,
$J^2=-{\bf 1}$.  In particular, the $SU(k-2)$ subgroup of the
$SP(k-2)$ global symmetry is anomaly free and can be identified
with a part of the standard model gauge group.\footnote{The full
$SP(k-2)$ can be gauged only with the inclusion of additional fields
in the fundamental of $SP(k-2)$ so as to avoid Witten's
global anomaly.}  If we wish to
embed color $SU(3)$ into the global symmetry, the minimal
size of the DSB gauge group is $SU(11)$.  If we wish to embed
the full $SU(5)$ extension of the SM gauge group, we need $SU(15)$.
Such large gauge groups result in an addition of 11 or 15 vector-like
color-triplet quarks to the standard model, respectively.  Note, on
the other hand, that the size of the supersymmetry breaking effects in
the mass spectrum is expected to be comparable to the scale of all
masses in these models because there is only one scale in the problem:
the scale parameter $\Lambda$ of the DSB gauge group.  Then this scale
must be around $10^{4}$~GeV in order to generate squark and slepton
masses of the desired size $\sim 10^{2}$--$10^{3}$~GeV, and as a
consequence the color gauge coupling constant blows up at scales
$5\times 10^{7}$~GeV, or $3\times 10^{6}$~GeV,
respectively.\footnote{The apparent blow-up of SM gauge
coupling constants may not necessarily be a disaster, if one can
regard such models as low-energy effective descriptions of other theories
which are ultraviolet safe.  Such a situation may arise if the
SM gauge groups are dual to asymptotically free gauge groups,
or if they are embedded into much larger groups which are
asymptotically free.  Unfortunately, no realistic example of this is
known.}

In principle one can solve this problem by employing a small coupling
constant: if one coupling constant in the superpotential is much less
than unity, the situation is similar to models with multiple scales.  Even
though there may be various ways in which a small coupling constant might
be obtained in a natural way \cite{AMM}, these models are not truly
satisfactory at this stage.

There have also been other attempts to simplify the structure of the OGM
models.  The idea has been to couple the DSB sector and vector-like
messenger fields directly in the superpotential, thereby eliminating
the messenger gauge field.  Unfortunately, this direction has not been
successful either.  One possibility was discussed by Intriligator and
Thomas \cite{IT1}, based on the vector-like DSB models \cite{IY,IT1}.
The vector-like messenger fields couple to the
O'Raifeartaigh-like singlet field in the DSB sector in the superpotential,
and obtain a supersymmetry breaking soft mass term.  However,
this model has a color-breaking
supersymmetric minimum of the potential, and the desired supersymmetry
breaking minimum is not absolutely stable.  Even if one accepts such an
unstable local minimum, a dynamical assumption is needed to ensure that the
singlet field develops an expectation value in its $A$-component, necessary
to generate the invariant mass for the messengers.  This region of
field space is strongly coupled, and there is no control over the K\"ahler
potential, so that it is impossible to tell whether or not such an expectation
value developes.  Another proposal by Hotta, Izawa and Yanagida
\cite{HIY} also tried to eliminate the messenger gauge field by coupling the
messenger fields to the vector-like model in the superpotential.
However, in order to break supersymmetry, they had to assume a non-generic
superpotential not justified by any symmetry, as well as making some
other dynamical assumptions.

\section{The scheme}
\label{sec:scheme}
\setcounter{footnote}{0}
\setcounter{equation}{0}

Here we describe our simple DSB scheme which maintains perturbative
gauge coupling unification by naturally generating a large invariant
mass for the fields that act as messengers, while
keeping $B\simeq 10^4$~GeV using non-renormalizable terms in the
superpotential.  The scheme we propose has three rather
ambitious goals: (1) the preservation of perturbative
gauge unification, (2) the natural supression of supergravity contributions
to the sfermion masses so as to maintain degeneracy, (3) no fine-tuning
of parameters or very small coupling constants.

In most models which break supersymmetry dynamically, there is a
non-perturbative superpotential which forces fields to move away from
the origin.  By lifting all flat directions at the classical level, by
adding suitable (renormalizable) terms to the superpotential, the
balance between the non-perturbative and tree-level terms determines a
stable minimum of the potential with a finite vacuum energy.

In contrast, suppose that the model has some flat directions $X$ which
are not lifted by the superpotential at the renormalizable level.
Then the vacuum runs away from the origin along such directions.  This
runaway behavior can be stopped by the possible addition of
non-renormalizable terms.  Therefore, the size of the field
expectation values along such flat directions, $\langle X \rangle$,
can be large while the vacuum energy, $\langle F_X \rangle^2$, stays
small.  On the other hand, such a large expectation value can give
large masses to other fields if they have renormalizable couplings to
$X$.  In particular it is possible that the fields which acquire
masses from $X$ act effectively as messengers.

As mentioned above, non-renormalizable models have already been
utilized by Poppitz and Trivedi \cite{PT2} to couple the DSB sector
directly to the SM.  For example, they considered a $SU(13)\times SU(11)$
model, which has an $SP(5)$ global symmetry into which a weakly gauged
$SU(5)$ can be embedded.  They found that this model allows a direct
coupling between the DSB sector and the SM gauge groups, while maintaining
the perturbative SM gauge coupling unification.  Unfortunately the models
suffered from the problem that both $\langle X \rangle$ and $\langle
F_{X}\rangle$ were too large, allowing the supergravity contribution
to the sfermion masses to dominate.\footnote{Actually, this problem
can be circumvented if one takes the scale of the higher dimension
operators to be less than the reduced Planck scale.  We return to this
point below eq. (4.5).}
Concretely, in their model,
vector-like fields are generated at a mass scale of $\sim \langle
F_{X} \rangle/\langle X \rangle$, which have both invariant mass and
supersymmetry breaking bilinear mass terms.\footnote{It was apparently
not appreciated in Ref.~\cite{PT2} that these heavy effective vector-like
fields can act as gauge mediation messengers.  Instead attention was
focused on the analogs of the light $b-\phi$ fields that we will
introduce in the next section.}  The gauge-mediated masses
for sfermions due to loops of $N$ vector-like messenger fields scale
as $\sim \sqrt{N} (\alpha/4\pi) \langle F_{X} \rangle/\langle X
\rangle$, and need to be around 100 to 1000~GeV.  A large $\langle X
\rangle$ thus implies a large $\langle F_X \rangle$.  On the other
hand, a large $\langle F_X \rangle$ generates supergravity
contributions to the scalar masses of order the gravitino mass,
$m_{3/2} \sim \langle F_X \rangle/M_*$.  To retain squark and slepton
degeneracy, which is the primary motivation for gauge mediation, one
needs to suppress $m_{3/2}$ to be at most 10\% of the
gauge-mediated contribution.\footnote{Or equivalently, 1\% in the
$({\rm mass})^2$ of the sfermions.}  This leads to an upper bound on
$\langle X \rangle$,
\begin{equation}
	\langle X \rangle \lsim 0.1 \sqrt{N} \frac{\alpha}{4\pi} M_*
		\simeq 2 \times 10^{15}~\mbox{GeV}
		\label{maxM}
\end{equation}
if we take $N \simeq 10$ and $\alpha \simeq 1/25$.  Their model gives
$\langle X \rangle \sim 6 \times 10^{16}$~GeV.  As we will show in the
following sections, our models satisfy this bound.

A general scheme for constructing a simple model of gauge mediation is
clear after these considerations.  Find a model where not all of the
classical flat directions are lifted at the renormalizable level.  The
model must have a relatively large global symmetry which can be
gauged.  Add suitable non-renormalizable terms to the superpotential
such that all flat directions are lifted.  Then the supersymmetry is
broken with a natural suppression of $\langle F_{X} \rangle \ll
\langle X \rangle^{2}$ so that the masses of extra fields coupled to
the standard model are heavy $\sim \langle X \rangle$ while keeping
the size of induced supersymmetry breaking small enough, at $\langle
F_{X} \rangle / \langle X \rangle \sim 10^{4}$~GeV.  A constraint here
is that the non-renormalizable operators should not be of too high a
dimension.  This was the reason the $SU(13)\times SU(11)$ model
failed in generating too large a supergravity contribution to the scalar
masses.  We need to keep the size of $A$-component expectation
value $\langle X \rangle$ to be within the range Eq.~(\ref{maxM}).

In fact, the phenomenological requirements can be conveniently
summarized in terms of the dimensionality of the operator which lifts
the relevant flat direction.  Suppose the flat direction $X$ is lifted
by an operator of mass dimension $n$, such that the non-renormalizable
operator in the superpotential is $W \sim \phi^n/M_*^{n-3}$ with $\phi$
standing generically for chiral superfields.  Then we obtain the
following crude estimate of the scales by requiring $B \equiv \langle
F_X \rangle/\langle X \rangle \simeq 10^4$~GeV,
\begin{eqnarray}
\langle X \rangle &\sim& M_* \left( \frac{B}{M_*} \right)^{1/(n-2)}, \\
\langle F_X \rangle &\sim& B M_* \left( \frac{B}{M_*} \right)^{1/(n-2)},
\end{eqnarray}
and
\begin{equation}
m_{3/2} \sim \frac{\langle F_X \rangle}{M_*}
	\sim B \left( \frac{B}{M_*} \right)^{1/(n-2)}.
\end{equation}
We would like to keep $\langle X \rangle$ large enough to maintain
perturbative unification.  In the one-loop renormalization group
analysis, one obtains $\langle X \rangle \gsim 10^{16-60/N}$~GeV for
$N$ extra ${\bf 5}+{\bf 5}^*$ pairs.  The constraint from a two-loop
analysis tends to be somewhat stronger than this.  On the other hand,
we need to keep $m_{3/2} \lsim 100$~GeV or so to have TeV-squarks
naturally degenerate at a percent level.  Therefore the dimensionality
of the operator should satisfy
\begin{equation}
2 + \frac{14.4}{2+60/N} \lsim n \lsim 9.2 \ .
\label{upper-lower-n}
\end{equation}
It should be noted that the precise constraints depend on
details of the models (such as an accidental cancellations, existence of
many operators with different dimensionalities, etc) and the upper and
lower bounds above are only ball-park numbers.  Furthermore, the
upper bound depends on the assumption that the non-renormalizable
operators are suppressed by the reduced Planck scale $M_*$.  If the
suppression is by a lower scale $M$, the upper bound becomes weaker,
$n\lsim 2 + (14.4 - \log_{10}(M_*/M))/(2-\log_{10}(M_*/M))$.

However, as we will discuss at length later, the particular model
which we will present in Section~5, as well as the models of
Poppitz and Trivedi, suffer from a serious problem.  The squark and
slepton masses are driven negative due to two-loop contributions to
their renormalization group evolution.  The origin of this problem is
clear.  Specifically, these contributions are due to fields charged
under the SM arising in the DSB sector, with masses of around 10~TeV,
and which in addition acquire large soft supersymmetry breaking
masses also at the 10~TeV scale.
We do not consider this a problem of the scheme itself.  If we
could find a model which does not produce light charged fields, the
problem can be trivially circumvented.
We believe that the scheme we discussed can be realized
while avoiding the problem of negative squark, slepton masses after
more exploration for models which break supersymmetry dynamically.

The recipe for the construction of a successful model can therefore
be easily summarized.  The DSB sector should ideally have the following
features.
\begin{enumerate}
\item It must accommodate a large global symmetry, such as $SU(5)$ or
at the very least $SU(3)$.\footnote{Even with a model $M$ which allows
only an $SU(3)$ global symmetry, gauge unification may be achieved
through triplicating the model as $M^3/Z_3$ and using trinification,
where the SM gauge group is embedded in $SU(3)^3/Z_3$.}

\item It leaves some flat
directions unlifted at the renormalizable level.

\item The addition of
non-renormalizable operators lifts all flat directions and the model breaks
supersymmetry dynamically.  The dimensionality of the non-renormalizable
operators should satisfy the constraint Eq.~(\ref{upper-lower-n}).

\item All directions with non-trivial quantum numbers under the standard
model gauge group are lifted at the renormalizable level to avoid light
charged fields which drive the sfermion masses negative via
two-loop running effects.
\end{enumerate}
Once one finds a model of DSB which satisfies the above criteria,
a direct coupling of the DSB sector and the SM gauge groups
is possible and achieves the three goals we desire.
Moreover, the model would not have a problem of negative sfermion
masses as will be described in Section~\ref{sec:fatal} and hence is
phenomenologically viable.  Finally, the model would have many
phenomenologically desirable features compared to the OGM models
as discussed in Section~\ref{sec:phenomenology}.

\section{A Model}
\label{sec:model}
\setcounter{footnote}{0}
\setcounter{equation}{0}

In this section, we describe an example of a model along the lines of the
scheme proposed in Section~\ref{sec:scheme}.
The model is based on an $SU(7)\times SU(6)$ gauge group.  It generates a
large $\langle X \rangle$ while naturally keeping $\langle F_X \rangle$
small, allowing perturbative gauge coupling constants up to the Planck scale.
In fact, the model has an $SU(5)$ global symmetry which can incorporate the
SM gauge group, so that the perturbative gauge unification in the
SSM is kept intact.  The model has certain aesthetically appealing features.
Its particle content is completely chiral, {\it i.e.}\/, none of the fields
are allowed to have mass terms.  Even though $M$ is high, the
supergravity contribution can be kept small enough for an appropriate
range of a coupling constant in the superpotential so that the squark
degeneracy is a natural consequence of the model.  However, we will show
later in Section~\ref{sec:fatal}, that when studied in detail
the model suffers from a fatal problem.  Although not
phenomenologically viable, the model in this section can be regarded
as a demonstration that the following requirements are simultaneously
achievable: (1) direct
coupling of the DSB sector and the standard model gauge group while
maintaining the perturbative unification of gauge coupling constants, (2)
enough suppression of the supergravity contribution, (3) no very small
coupling constant.  We now describe this model in detail.


The particle content of the model is quite simple.  It is the same as the
$SU(N)\times SU(N-1)$ models proposed in Ref.~\cite{PST1} except for the
addition of a field $\phi$.  Under the $SU(7)\times SU(6)$ gauge group, we
introduce three sets of fields, $Q({\bf 7}, {\bf 6})$, $L_{I}(\bar{\bf
7}, {\bf 1})$ for $I=1,\cdots,6$, and $R^{I}({\bf 1}, \bar{\bf 6})$
for $I = 1, \cdots 7$.  We distinguish the first five of $L_{i}$ and
$R^{i}$ ($i=1,\cdots, 5$) from the rest ($L_{6}$, $R^{6}$, and
$R^{7}$ -- in the general case $L_{6},...,L_{N-1}$ and $R^{6},...,R^{N}$)
because we would like to impose an $SU(5)$ global symmetry
which can be gauged.  We also need a field $\phi^{i}$ ($i=1,\cdots,5$)
which is a singlet under the $SU(7)\times SU(6)$ gauge group, but transforms
as a fundamental under $SU(5)$, in order to cancel the
$SU(5)^{3}$ anomaly.  The addition of this field implies that the
flat-direction analysis of \cite{PST2} must be redone, and it is a
non-trivial fact that quantum mechanically this model still dynamically
breaks supersymmetry, as we will see explicitly below.

The most general superpotential compatible with the $U(1)$ and
$U(1)_{R}$ symmetry listed in  Table~1 (these two $U(1)$'s are
non-anomalous) is given by
\begin{equation}
	W = \lambda L_{i} Q R^{i} + \lambda' L_{6} Q R^{6}
		+ \frac{g}{M_{*}} L_{i} Q R^{6} \phi^{i}
		+ \frac{\alpha}{M_{*}^{3}} b_{6}
		+ \frac{h}{M_{*}^{4}} b_{i} \phi^{i} .
		\label{W}
\end{equation}
Here and below, the ``baryon'' operators $b_{I}$ are defined by
\begin{equation}
b_I = {1 \over {6!}}\epsilon_{J I_{1} \cdots I_{6}}
\epsilon^{\alpha_1 \cdots \alpha_6}
R^{I_{1}}_{\alpha_1} \cdots R^{I_{6}}_{\alpha_6}
\end{equation}
We chose the scale of non-renormalizable operators to be the reduced
Planck scale, $M_*$.  This is the worst choice from the point of
view of suppressing the supergravity contribution as we will see later.
Still, the model suppresses the supergravity contribution enough because
the dimensionality of the operator is $n=6$ satisfying the constraint
Eq.~(\ref{upper-lower-n}) discussed in Section~\ref{sec:scheme}.

\begin{table}
\caption[1]{The charge assignments of the fields in the
$SU(7)\times SU(6)$
model under the non-anomalous $U(1)$, $U(1)_R$, and $SU(5)$ symmetries.}
\begin{displaymath}
\begin{array}{cccccccc}
& Q & L_i & L_6 & R^i & R^6 & R^7 & \phi \\ \hline
\mbox{U(1)} & +5 & -12 & +30 & +7 & -35 & -35 & +42\\
\mbox{U(1)}_R & -2 & +4 & -10 & 0 & +14 & +2 & -14\\
\mbox{SU(5)}&1&{\bf \bar{5}}&1&{\bf 5}&1&1&{\bf 5}
\end{array}
\end{displaymath}
\end{table}


Here we summarize the main points of the analysis, whose details can be
found in the appendix. The $D$ flat directions of the theory are
parameterized by the gauge invariant operators $b_{I}$,$Y^{iI} = L^{i}Q R^{I}$,
and ${\cal B}=\mbox{det}(LQ)$ with a constraint:
$\epsilon_{i_{1}i_{2}\cdots i_{6}}Y^{i_{1}I_{1}}\cdots
Y^{i_{6}I_{6}} \propto {\cal B} \epsilon^{I_{1}I_{2}\cdots
I_{6}I_{7}}b_{I_{7}}$.  In the appendix we show that our superpotential
forces all these operators to vanish classically due to the conditions
for a supersymmetric vacuum, and hence all these D flat directions are
lifted at the classical level.  It is non-trivial that even after adding
the singlet field $\phi$, all classical flat directions except for
$\phi$ are lifted.  We will see later that the quantum effects lift the
origin. Since the $b_{I}$ directions are lifted only by
non-renormalizable operators, the fields roll down along one of these
directions.  Therefore, it is useful for later purposes to analyze the
classical Lagrangian along these directions in the absence of
non-renormalizable terms in the superpotential.

Along the $b_{7}$ direction, $R^{1}$ to $R^{6}$ acquire the
same expectation values $\rho$,
\begin{equation}
	(R^{1}, R^{2}, R^{3}, R^{4}, R^{5}, R^{6}, R^{7})
		= \left( \begin{array}{ccccccc}
			\rho&0&0&0&0&0&0 \\
			0&\rho&0&0&0&0&0 \\
			0&0&\rho&0&0&0&0 \\
			0&0&0&\rho&0&0&0 \\
			0&0&0&0&\rho&0&0 \\
			0&0&0&0&0&\rho&0 \end{array} \right) .
\end{equation}
This configuration breaks the $SU(6)$ gauge group completely.
All components of $Q$ and $L$ fields are massive along $b_{7}$
direction, and can be integrated
out, leaving an unbroken pure $SU(7)$ theory.  It is
this group which generates a non-perturbative superpotential in
quantum analysis.  The situation is the same for arbitrary $b_{I}$
direction as long as the determinant of $Q$, $L$ mass matrix is
non-vanishing.
The K\"{a}hler potential for $b_{I}$ at the classical level can be
obtained using the method of Poppitz and Randall \cite{PR}.
The calculation is described in the appendix, and we find
\begin{equation}
	K = 6 (b^{*J} b_{J})^{1/6}.
\end{equation}

At the quantum level, the effective pure $SU(7)$ gauge coupling depends on
$\langle R \rangle$ through matching at the masses of the $Q,L$ fields.
Gaugino condensation in the $SU(7)$ gauge group generates a
non-perturbative superpotential for $R$ which prefers larger
$\langle R \rangle$.  The balance between this non-perturbative term and
tree-level term $b_6$ determines the value of $\langle R \rangle$ as
well as its $F$-component.  Now, a non-vanishing $F$-component of $R$
implies that the massive $Q$ and $L$ fields also have supersymmetry breaking
bilinear mass terms.  We will find that only $b_{6,7}$ acquire vacuum
expectation values, so an $SU(5)$ symmetry is left unbroken
by the dynamics, corresponding
to a vacuum expectation value for the 7$\times$6 $R$ matrix
which is proportional to the identity in the upper 5$\times$5
block. The unbroken $SU(5)$ is the diagonal subgroup of the
original global $SU(5)$ and the gauged $SU(6)$ symmetry.
The $Q,L$ fields then contain 7 pairs of
$({\bf 5} + {\bf 5^*})$ under the $SU(5)$,
with supersymmetric masses $\langle R \rangle
\equiv \langle R^i_i \rangle$ and supersymmetry breaking bilinear mass terms
$\langle F_R \rangle \equiv \langle F_{R^i_i} \rangle$.
The $Q,L$ fields, then, mediate supersymmetry breaking effects to
the standard model at order $7 (\alpha/4\pi) \langle F_R
\rangle/\langle R \rangle$.  Below we briefly describe
the analysis of the model; details are given in the appendix.

Explicitly, the SU(7) gaugino condensate generates a
non-perturbative superpotential,
\begin{equation}
	W_{\rm non-pert} =
		(\Lambda^{15} (\mbox{det}' \lambda) b_{7})^{1/7} .
\end{equation}
which prefers larger $b_{7}$.  Here, $\mbox{det}'$ refers to the
determinant of the $\lambda_{iI}$ matrix for $I=1,\cdots,6$.
Therefore, it is consistent to analyze the model along $b_{I}$ flat
directions where $SU(6)$ is completely broken, and a perturbative
K\"{a}hler potential for $b_{I}$ is valid.

The complete Lagrangian far along the $b_I$ flat directions
is then given by
\begin{eqnarray}
	K &=& 6 (b^{*J} b_{J})^{1/6} + \phi^*_i \phi^i, \\
	W &=& (\Lambda^{15} (\mbox{det}' \lambda) b_{7})^{1/7} +
		{\alpha \over {M_{*}^3}}b_{6} +  {h \over {M_{*}^4}} b_{i} \phi^{i}.
\end{eqnarray}
Following the analysis presented in appendix, a numerical minimization
of the potential for the case $N=7$ gives the location of the minimum
\begin{eqnarray}
	b_{N-1} &= & -0.0702 \alpha^{-7/6} \Lambda^{5/2}
		M_{*}^{7/2} ,\\
	b_{N}   &=& 0.0791 \alpha^{-7/6} \Lambda^{5/2}
		M_{*}^{7/2} .
\end{eqnarray}
Here and below, we absorb the unimportant factor $\mbox{det}'$ into the
definition of $\Lambda$.

Now we can discuss the various
mass scales in the model and give numerical results.
We will show that the supergravity contribution can be
suppressed enough compared to the gauge-mediation contribution if
$\alpha$ is not too small.  We find that the mass of the
anomaly-cancelling field $\phi$ tends to be light.  However it
can be beyond the experimental lower bound with a somewhat small but not
unnatural value of $\alpha$.  There is a parameter region which is
consistent with both requirements.

First of all, the scale of the vacuum expectation value of
$A$- and $F$-components of the $R^k_k$ field, which generates
the masses and supersymmetry breaking for the $Q,L$ messenger fields
charged under $SU(5)$, is (see appendix)
\begin{eqnarray}
& &	\langle R \rangle =.688 \alpha^{-7/36} \Lambda^{5/12} M_{*}^{7/12} \\
& & \langle F_{R} \rangle =.0425 \alpha^{1/36} \Lambda^{2}
	\left(\frac{\Lambda}{M_{*}}\right)^{1/12}
\end{eqnarray}
In order to get correct mass scales for the gaugino masses, we
require
\begin{equation}
	B \equiv \frac{\langle F_{R} \rangle}{\langle R \rangle}
	\simeq 10^{4}~\mbox{GeV} ,
\end{equation}
which generates the gluino mass of 600~GeV.
By setting $M_{*} = 2.4 \times 10^{18}$~GeV, we obtain
\begin{eqnarray}
	\Lambda = \alpha^{-2/15} \ 3\times 10^{10}~\mbox{GeV}
	  \\
	\langle F_{R} \rangle
        =\alpha^{-1/4} \ (3\times 10^{9}~\mbox{GeV})^{2}
	\\
	\langle R \rangle
	= \alpha^{-1/4} \ 8\times 10^{14}~\mbox{GeV}
\end{eqnarray}
Note that $\langle R \rangle$ is large enough so that the gauge
coupling constants stay completely perturbative up to the reduced
Planck scale even though we have added effectively seven {\bf
5}$+${\bf 5}$^{*}$ pairs.

There is a light $b_{k}$-$\phi^{k}$ fermion since its mass is
generated by the highest dimension operator in the model.  Its mass is
given by
\begin{equation}
	m_{\phi}^{2} = 0.0237 V_{0} \frac{h^{2}}{\alpha^{2} M_{*}^{2}}
		= (\alpha^{-5/4} h\times 13~\mbox{GeV})^{2}.
\end{equation}
Phenomenology requires such extra vector-like matter must be heavier
than $\gsim 200$~GeV for quark-like and $\gsim 80$~GeV for
lepton-like fields.\footnote{These numbers are by no means precise.}
Barring the fact that the invariant masses are enhanced due to the
renormalization group running, we require the original $m_{\phi}$ to
be larger than 100~GeV which is actually a too strong requirement.  This
bound translates into $\alpha < 0.19$ for $h = 1$.  Therefore, the
phenomenological constraint is satisfied without taking an unnaturally
small coupling constant.

The supergravity effect scales as $m_{3/2}^{2} = \langle
V \rangle / 3 M_{*}^{2} = 0.0234 \alpha^{1/18} \Lambda^{2}
(\Lambda/M_{*})^{13/6}$ where $\langle V \rangle$ is the vacuum energy after
supersymmetry breaking. Numerically, we find this is
$m_{3/2} = \alpha^{-1/4}\times 13~\mbox{GeV}$.
In order to keep the squark degeneracy at the level of 1\%, we need
$m_{3/2} \lsim 10^{2}$~GeV for squarks, or $\alpha > 3 \times
10^{-4}$.  Even
with a more stringent constraint $m_{3/2} \lsim 30$~GeV, still
$\alpha$ as small as $0.04$ is allowed.  We should also emphasize
that choosing $M_{*}$ to be the reduced Planck scale makes the supergravity
effect as competitive as possible to the gauge mediated effect. Any smaller value
for $M_{*}$ makes the supergravity contribution further negligible.
Therefore, for a natural choice of $3\times 10^{-4} \lsim \alpha \lsim
0.19$, we obtain a heavy enough $b$-$\phi$ fermion while suppressing
the supergravity contribution not to spoil the squark degeneracy more
than a percent level.

In summary, we have demonstrated that we can simultaneously achieve our
three goals in this model.  However, as we have mentioned
already, the model faces a new problem, which is directly tied
to the existence of the relatively
light $b-\phi$ multiplets.  Before discussing this serious problem,
however, we wish to outline, in a general way, the phenomenological
advantages of a scheme in which both the mass scale of the messengers,
and the fundamental scale of supersymmetry breaking are higher than
in the OGM models.

\section{Phenomenology}
\label{sec:phenomenology}
\setcounter{footnote}{0}
\setcounter{equation}{0}

We now discuss general phenomenological features distinguishing
our scheme from  the original models.  The only ingredient used in
this section is a high $\langle X\rangle$ with $\langle F_{X}
\rangle/\langle X \rangle \sim 10^{4}$~GeV.  Specifics could
easily depend on details of individual models, but the features we
discuss in this section are generic to any models which follow our
scheme.  Most of the problems mentioned in section~\ref{sec:original}
are improved upon.  The only point which is somewhat worse than
in the OGM models is the cosmological problem associated with
Polonyi-like fields; however it is not as serious as in the hidden sector
models.

\begin{itemize}
\item {\bf Superparticle mass spectrum and electroweak symmetry
breaking.}

In our scheme, the mass scale of effective messengers is rather high
in order to keep the gauge coupling constants perturbative.  This simple fact
provides us with a distinct superparticle mass spectrum differing from
the OGM models \cite{DN,DNS,DNNS}.  The masses tend to be much closer
to each other.  In particular, the splitting between squarks and
sleptons are much smaller \cite{messenger-scale}.

This point has two phenomenological consequences.  One is that one can
tell two schemes apart by measuring superparticle masses precisely.  In
fact, future colliders will have the  capability of measuring superparticle
masses quite well; to the 10\% level for squarks and gluino and
to the 1\% level for the mass splitting between the first and second
neutralinos at the LHC, depending upon certain theoretical assumptions
\cite{Ian}, and to the 1\% level for any
superparticles without any assumptions at an $e^+ e^-$ linear collider
\cite{Tsukamoto,FF,FPMT,ZDR}.  Another point is that the fine-tuning in
electroweak symmetry breaking becomes mild.  One of the problems in the
original scheme \cite{DN,DNS,DNNS} is that the squarks are much heavier
than the sleptons, and the negative stop contribution to the  Higgs mass
squared is too large, forcing an order 1\% fine-tuning so as to
correctly achieve electroweak symmetry breaking \cite{ACHM}.  On
the other hand, the mass splitting between the squarks and sleptons is
much smaller in our scheme and hence the fine-tuning is less severe; we
do not expect a fine tuning at more than the 10\% level.

\item  {\bf $\mu$-problem}

The $\mu$-problem can be solved in a simple way by extending the
minimal particle content to include a singlet (NMSSM) with the
following superpotential,
\begin{equation}
	W = \lambda_{1} H_{u} H_{d} S + \lambda_{2} S^{3} .
\end{equation}
This simple model does not work in the OGM models because it posses an
approximate $U(1)_{R}$ symmetry which is broken only by small
soft supersymmetry breaking trilinear couplings
of order 10~GeV. A spontaneous breaking of such an
approximate global symmetry leads to an unacceptably light axion-like
pseudo-scalar Higgs boson which appears in $Z$-decay in  combination
with a light neutral Higgs boson \cite{DN}.  In our scheme, however, the
trilinear couplings are induced from gaugino masses with a large
logarithm and are much larger.

It is known that the coupling of a singlet to $H_u H_d$ may destabilize
the hierarchy in the hidden sector models.  If $H_u$ and $H_d$ are
embedded into multiplets unified with massive color-triplets, their loops
may induce a tadpole for the singlet of order
$V \sim (g^2/16\pi^2) (\ln M_{H_C}^2/m_Z^2) M_{H_C} m_{3/2}^2 S$.
In our case the problem is somewhat improved
because the soft supersymmetry breaking is power-suppressed for the
color-triplet Higgs.  However it can still be too large.  The
soft supersymmetry breaking for the color-triplets is of order $\sim
(g^2/16\pi^2)^2 (F_X/M_{H_C})^2$, and the tadpole is given by
\begin{equation}
V \sim
\left(\frac{g^2}{16\pi^2}\right)^3
\left(\ln \frac{M_{H_C}^2}{m_Z^2}\right)
M_{H_C} \left(\frac{F_X}{M_{H_C}}\right)^2 S.
\end{equation}
Since we fix $F_X/X \sim 10^4$~GeV, this is problematic if
$X \gsim 10^9$~GeV.  Fortunately, the existence of this tadpole is
model-dependent and certain theories do not
produce it.  In general, the models which naturally avoid proton decay
via $H_C$ exchange do not produce this tadpole.  Examples are
the flipped SU(5) model \cite{flipped}, and the models with Babu--Barr
mechanism \cite{Babu-Barr}.  The tadpole problem certainly puts
constraints on GUT-model building but appears surmountable.

\item  {\bf Exotic stable particles}

Possible stable particles from the DSB sector are mostly much heavier
$\sim \langle X \rangle$ than in the OGM models.  Therefore,
primordial inflation could well dilute them away.  There typically are
particles around $10^4$~GeV scale which correspond to flat directions
lifted only by higher dimension operators.  As will be discussed later
in Section~\ref{sec:fatal}, we would like such ``light'' particles to
transform trivially under the standard model not to drive squark and
slepton masses negative at the weak scale.  Also by definition, these
fields have expectation values and hence there are presumably no
conserved quantum numbers associated with them.  Therefore, they are
likely to decay and there is no problem with the closure limit.
However, their scalar components act similarly as the Polonyi field in
the hidden sector scenario, and hence might have a coherent oscillation.
We will come back to this question shortly.

\item  {\bf $R$-axion}

The decay constant is much higher than in the OGM models.  It is
interesting that one may have a decay constant in an interesting range,
$\langle X \rangle \sim 10^8$--$10^{13}$~GeV, such that the $R$-axion
may be a viable candidate for the QCD axion.  As discussed in Section
~\ref{sec:scheme}, the decay constant has to be less than about $\langle
X \rangle \lsim 2\times 10^{15}$~GeV in order to suppress the supergravity
contributions to the squark and slepton masses.  For a decay constant above
$10^{13}$~GeV, the coherent oscillation of the axion may overclose the
Universe.  However, a decay of long-lived particle may dilute the axion
coherent oscillation, and the axion decay constant up to $10^{15}$~GeV
may be allowed \cite{KMY}.  It is noteworthy that the same range
of $\langle X \rangle \lsim 10^{15}$~GeV is preferred both by viable
cosmology and the suppression of supergravity contributions to the squark
and slepton masses.

In order for the $R$-axion to be the QCD axion, one needs to suppress
the possible higher dimension operators which could break $R$-symmetry
explicitly \cite{Marc}.  This still remains as a significant
constraint on Planck-scale physics.  It is worth  pointing out that the
$R$-axion originates from gauge non-singlet fields in these models so that
it is somewhat easier to forbid operators up to certain dimensionalities as
an accidental consequence of gauge symmetries which are stable against
quantum gravitational effects.

\item  {\bf Gravitino problem.}

In the OGM models, the gravitino mass is expected to be order
100~keV. In this mass range, the decay of usual LSP's into the
gravitino overcloses the Universe \cite{MMY,Andre}.  One should
suppress the temperature so that most of the SUSY particles were never
created.  In our scheme, the gravitino mass is higher,
and the upper bound on the reheating temperature is
much weaker.  However, a too large gravitino mass $m_{3/2} \geq 10$~GeV is
forbidden because the decay of LSP's into gravitino occurs too late,
and destroys the success of nucleosynthesis.  It is interesting that
this constraint is similar to the other constraint to suppress the
flavor-non-universal SUGRA contribution to the scalar masses.  They
are consistent with each other.

\item {\bf $ll\gamma\gamma$ events.}

The gravitino decay constant in our scheme is much higher and it is
impossible to have the lightest superparticle in the SSM decay
into the gravitino inside a typical collider detector.
We do not except $ll\gamma\gamma$-type events arising from
pair production of sleptons each decaying into a gravitino, a lepton and
a photon.  Overall the collider phenomenology is somewhat similar
to the hidden sector case except the following point.
It is not a cosmological problem for the lightest superparticle in the
SSM to be a charged particle because it decays into a gravitino
well before nucleosynthesis.  Therefore, a light charged
superparticle, such as a slepton or a chargino, may be a viable
lightest SSM superparticle and charged tracks insider the
detector may be a signal for this scheme of mediation.

%
%

\item {\bf String Moduli.}

The moduli in the string theory acquire masses only through
supersymmetry breaking, and therefore their masses are expected to be
of the same order as the gravitino mass $m_{3/2}$.  In the OGM models,
this is as small as 100~keV, and hence the moduli are stable.  Their coherent
oscillations, with initial values of order Planck scale as expected
generically for moduli fields, overclose the Universe by 15 orders of
magnitude \cite{Andre}.  Since the mass scale is so low, it is quite
difficult to eliminate them.  In our scheme, the moduli masses are
much higher. Even though their decays would occur after
nucleosynthesis and would cause disastrous effects if there
weren't any dilutions, a thermal inflation \cite{LS} could easily
eliminate them.

\item {\bf Pseudo-Polonyi Problem.}

Models in our scheme generically produce particles around the
$10^{4}$~GeV scale.  The particles correspond to directions in the
field space which are not lifted by the renormalizable superpotential,
but are lifted by non-renormalizable terms.  Since the size of the
potential energy is $\sim \langle F_{X} \rangle^{2}$ while the natural
size of the field variation is $\sim \langle X \rangle$, the mass
generically turns out to be order $m_{X}^{2} \sim \langle F_{X}
\rangle^{2} / \langle X \rangle^{2} \sim (10^{4}~\mbox{GeV})^{2}$.
Because of its large vev, however, the field interacts with other
fields via interactions suppressed by the scale $\langle X \rangle$.
As a result, its decay rate is quite suppressed.
The precise expression for its decay rate is model-dependent.  For
instance, the baryon-fields $b_6$ and $b_7$ in the $SU(7)\times SU(6)$
model are the Polonyi-like fields in this case.  Their decay
proceeds via a loop diagram of the heavy $Q$-$L$ multiplets into the SM
gauge multiplets, {\it e.g.}\/, two gluons or gluinos.  For both of them,
the decay rate is suppressed further by the loop factor with 7
color-triplets, $(7 \alpha_{s}/\pi)^{2}$.  In general, we expect the
decay rate for a Polonyi-like field to be
\begin{equation}
	\Gamma_{X} \sim \frac{1}{8\pi}
		\left( \frac{N \alpha_{s}}{\pi} \right)^{2}
		\frac{m_{X}^{3}}{\langle X \rangle^{2}}
		\sim (10^{-2}~\mbox{sec})^{-1}
		\left( \frac{m}{10^{4}~\mbox{GeV}} \right)^{3}
		\left( \frac{\langle X \rangle}{10^{15}~\mbox{GeV}}
			\right)^{-2} N^{2}.
\end{equation}
Here, $N$ is the number of multiplets which contribute to the loop
diagram.

If such a long-lived scalar particle has an initial field amplitude
of order $\langle X \rangle$, its coherent oscillation
acquires a large energy density and its subsequent decay produces
an enormous energy and entropy.  In the hidden sector scenario of
supersymmetry breaking, this decay tends to occur after
nucleosynthesis, and destroys the success of the big-bang nucleosynthesis
predictions for the  light element abundances.  Fortunately in our case, we
need $\langle X \rangle \lsim 2 \times 10^{15}$~GeV (\ref{maxM}) to
suppress supergravity contribution to the scalar masses, and hence the
decay of the field $X$ is likely to occur before the nucleosynthesis
time, $\tau \sim 1$~sec, even for the worst case, $N = 1$.
This difference is due to a less
suppressed coupling ($\langle X \rangle$ vs $M_{*}$) and a larger
mass ($m_{X} \sim 10^{4}$~GeV vs $m_{3/2} \sim 100$~GeV).

The entropy production due to the decay of the coherent oscillation may
still be a concern in general.  The dilution factor is given by
$D \sim \sqrt{8\pi} \langle X \rangle^3 \pi / (N \alpha_{s} m_X M_*^2)
\sim 3 \times 10^{16} (\langle X
\rangle/M_{*})^{3}$, and hence not very important for $\langle X \rangle
\lsim 10^{13}$~GeV.  For a higher $\langle X \rangle$, the entropy
production may become more significant: for a possibly maximum $\langle
X \rangle \sim 2 \times 10^{15}$~GeV required from suppressing
supergravity contribution, $D \lsim 10^7$.  It is never as bad as in the
hidden sector case $D \sim \sqrt{8\pi} M_*/m_{3/2} \sim 10^{16}$.  An
efficient baryogenesis such as in Affleck--Dine scenario \cite{AD} could
well be sufficient.

\end{itemize}

\section{A New Problem}
\label{sec:fatal}
\setcounter{footnote}{0}
\setcounter{equation}{0}

We have seen in Section~\ref{sec:model} that one can achieve the three
goals of perturbative unification despite the direct coupling, a natural
suppression of supergravity contribution, and no need for a fine-tuning
or a very small coupling constant, simultaneously along the lines of the
scheme we described in Section~\ref{sec:scheme}.
However, the model
presented in the Section~\ref{sec:model} suffers from a fatal flaw: the
standard model gauge group gets broken in running down from the high
scale to the weak scale.  This problem has not been discussed in the
literature, and we will describe it in detail in this section.  It has
to be emphasized that the problem is model-dependent and may  not exist
in certain other models with directly coupled DSB sector.  We
are currently not aware of any DSB models which achieve all three goals
without the problem described in this section.  Nonetheless it seems
likely that a viable model in our scheme can be found.

The problem is that the model in Section~\ref{sec:model}
has light chiral multiplets $b$, charged under the standard model, with
supersymmetry breaking in the form of soft scalar masses of order $m_b^2
\sim (\langle F_R \rangle/\langle R\rangle )^2$.  While these multiplets
only communicate to the ordinary sector via gauge interactions, there is
a negative, logarithmically divergent two loop contribution to the
running of the ordinary sfermion soft masses due to the soft mass of
$b$, which wins over the positive contribution from the $Q,L$ messengers
and the positive contribution from 1 loop running due to the standard
model gaugino masses.  In the absence of the $\phi$ field, the
fermionic components of
the $b_i$ superfields are the massless fermions required by 't Hooft
anomaly matching, and they pick up a mass of order the weak scale after
coupling to $\phi$. The scalar components of the $b_i$ superfield,
however, have a soft scalar mass of order $(\langle F_R \rangle /
\langle R\rangle)^2$.  Numerically, we find that $(\langle F_R \rangle /
\langle R\rangle)^2=.0038 \Lambda^{10/3} M^{-4/3}$ while $m_{b_i}^2 =
.104 \Lambda^{10/3} M^{-4/3}$, so $m_{b_i}^2 \sim 27 (\langle F_R
\rangle / \langle R\rangle)^2$.  Consider now the renormalization of the
squark masses from the scale $v$ to lower scales.  The relevant RGE
keeping only the strong gauge coupling is \footnote{The 1-loop
hypercharge $D$ term contribution from $m_{b_i}^2$ vanishes since the
$b_i$ scalar masses respect $SU(5)$ invariance and so Tr$Y m_{b_i}^2 =
0$.}  $$ {d\over{d t}} m_{\widetilde{q}}^2 = -{1\over{(16
\pi^2)}}{32\over3} g_3^2 M_3^2 + {1\over{(16 \pi^2)^2}} {16\over3} g_3^4
m_{b_i}^2 $$ where the first term is the positive 1-loop contribution
from the gluino mass $M_3$, the second is the negative two loop
contribution from $m_{b_i}^2$, and we have neglected the two loop gluino
mass contributions.  We use the notation $t = \ln \mu$ here and below.
We can treat $m_{b_i}^2$ as constant since all the contributions to its
running from the gluino and sfermion masses are negligible compared to
$m_{b_i}^2$. Also, we can treat $M_3/g_3^2$ as fixed at its initial
value coming from the $Q,L$ messengers $M_3/g_3^2 = N/(16 \pi^2)
(\langle F_R \rangle/\langle R\rangle )$ with $N=7$.  Then, we find
\begin{equation}
{d\over{d t}} m_{\widetilde{q}}^2 =
	-{16\over3}{1\over{(16 \pi^2)^2}} g_3^4
	\left(\frac{\langle F_R \rangle}{\langle R\rangle} \right)^2
\left({2 N^2 g_3^2 \over {16 \pi^2}} - 27\right)
\label{N2vs27}
\end{equation}
For $N=7$, the quantity in the brackets gives a negative
contribution to the squark masses, and for running from
$\langle R \rangle \sim 10^{14}$ GeV to the weak scale, we have
\begin{equation}
\Delta m_{\widetilde{q}}^2 \sim
	-0.1 \left(\frac{\langle F_R \rangle}{\langle R\rangle} \right)^2
\end{equation}
which dominates over the positive contribution from the $Q,L$ messengers
\begin{equation}
m_{\widetilde{q}}^2(\langle R \rangle)={16 N \over 3}{1\over {(16 \pi^2)^2}}
\left(\frac{\langle F_R \rangle}{\langle R\rangle} \right)^2
\sim .002 \left(\frac{\langle F_R \rangle}{\langle R\rangle} \right)^2.
\end{equation}

The most stringent constraints on the relative size of $(\langle
F_R\rangle/\langle R\rangle)^2$ and
$m_{b_i}^2$ actually arises from the renormalization
group flow of the right-handed and left-handed SSM sleptons.
The renormalization group equations, including the
dangerous two-loop contributions of the light $b_k$ scalars, are:
\begin{eqnarray}
{d\over{d t}} m_{\widetilde{L}}^2 &=&
-{1\over{(16 \pi^2)}}\left( 6 g_2^2|M_2|^2 +
\frac{6}{5} g_1^2 |M_1|^2 \right) \nonumber \\
& &
+ {1\over (16\pi^2)^2}
\left( 3g_2^4 m_b^2 + \frac{3}{5} g_1^4 m_b^2
\pm \frac{6}{5} g_1^2 m_b^2(\frac{8}{3}g_3^2-
\frac{3}{2}g_2^2 -\frac{1}{6}g_1^2) \right)
\label{eq:lhdRGE}
\end{eqnarray}
for the left-handed slepton doublet, and
\begin{eqnarray}
{d\over{d t}} m_{\tilde{l}}^2 &=&
-{1\over{(16 \pi^2)}}\left(\frac{24}{5} g_1^2 |M_1|^2 \right)
\nonumber \\
& &
+ {1\over (16\pi^2)^2}\left( \frac{12}{5} g_1^4 m_b^2
\mp \frac{12}{5} g_1^2 m_b^2(\frac{8}{3}g_3^2-
\frac{3}{2}g_2^2 -\frac{1}{6}g_1^2) \right)
\label{eq:rhdRGE}
\end{eqnarray}
for the right-handed sleptons.  The $\pm$ signs in the two-loop
terms of Eqs.~(\ref{eq:lhdRGE},\ref{eq:rhdRGE}) correspond to the
assignment of $b_k$ to either a $5$ or $\bar 5$ of SU(5)
respectively, and arise from $D$-term interactions.  (Note that
we are neglecting two-loop contributions to these RG equations
proportional to the soft mass-squared of $\phi$, since these soft
masses are suppressed relative to those of the $b_k$ states.)
The renormalization group equation for the right-handed slepton mass
potentially provides the severest constraint on the relative
size of $(\langle F_R\rangle/\langle R\rangle)^2$ and $m_b^2$, since the
term of ${\cal O}(g_1^2 g_3^2 m_b^2)$ dominates the gaugino
contribution (which varies as ${\cal O}(g_1^6 N^2 (\langle
F_R\rangle/\langle R\rangle)^2)$)
to the greatest extent.  However, if we use
the freedom in assigning the $b_k$ states to either the $5$ or
$\bar 5$, to choose the negative sign in Eq.~(\ref{eq:rhdRGE})
-- the $5$ assignment --
then the right-handed slepton mass-squared is actually increased
by the potentially dangerous two-loop terms.
The sign of the two-loop terms for the left-handed slepton
doublet is then fixed, and this state is the first to have
its mass-squared driven negative by RG flow.

The difficulty above is not specific to the $SU(7)\times SU(6)$ model
but is quite generic: we always expect some of the moduli to gain soft
masses of order $(\langle F_R\rangle/\langle R\rangle)^2$ in any theory
with a single scale where some of the flat directions are lifted by high
dimension operators.  If these fields are charged under the standard
model, the two loop contribution to the ordinary sfermion masses are
problematically large and negative.  Note, however, that the sign of
the contribution to ordinary sfermion masses depends crucially on $N$
and the relative enhancement or suppression of the charged modulus field
mass relative to $(\langle F_R\rangle/\langle R\rangle)^2$.  For
instance, the situation is worsened in the $SU(7)\times SU(6)$ model,
since $m_{b_i}^2$ was enhanced by a factor $\sim 27$ relative to
$(\langle F_R\rangle/\langle R\rangle)^2$.  If there was a suppression
rather than an enhancement, the positive gaugino contribution could have
dominated and there would be no problem.

Despite the impression from Eq.~(\ref{N2vs27}), a larger $N$ or a lower
scale $M < M_{*}$ of the non-renormalizable operators does not solve the
problem.  To see whether such a flaw occurs, we need to calculate the
exact expressions for $(\langle F_R\rangle/\langle R\rangle)^2$ and
$m_b^2$ in terms of $\Lambda$, $M$ (the scale of the non-renormalizable
operators, no longer restricted to be $M_*$), and $N$, for a general
model.  As discussed in the appendix, the formula for
$(\langle F_R\rangle/\langle R\rangle )$ is given by
\begin{equation}
\frac{\langle F_R\rangle}{\langle R\rangle}
	= m {(\beta_{N-1}+\beta_N^{1/N}/N) \over
	(\beta^\dagger \beta)^{1/(N-1)}},
\label{eq:fv}
\end{equation}
with
\begin{equation}
m =
\alpha^{(N+1)/(N-1)^2} \Lambda^{(2N+1)(N-3)/(N-1)^2}
M^{-(N-4)(N+1)/(N-1)^2},
\end{equation}
while the mass of the $b$-scalars are given by
\begin{equation}
m_b^2 = m^2 (\beta^\dagger \beta)^{(N-3)/(N-1)}
    \frac{N-2}{N-1} \left(
        1 + {1\over N^2} (\beta_N^\dagger \beta_N)^{-(N-1)/N}
        - \frac{1}{\beta^\dagger \beta} |\beta_{N-1}+\beta_N^{1/N}/N|^2
        \right).
\label{eq:mb2}
\end{equation}
The subleading contributions to $m_b^2$ arise from the diagonalization
of the $b-\phi$ system, and the last $\lambda$-dependent
term of the full scalar potential.  Neither of these additional
complications change our conclusions.

Equations (\ref{eq:fv}) and (\ref{eq:mb2}) show that both $m_b^2$ and
$(\langle F_R\rangle/\langle R\rangle)^2$ depend on the quantity $m^2$,
and thus their dependence on all dimensionful parameters ($\Lambda$ and
$M$, and therefore the coupling $\alpha$ as well) is identical.  Thus we
gain nothing by working at fixed $N$ but allowing $\Lambda$ and $M$ to
vary.  However, since the SSM gaugino masses vary as $N\alpha_i
(\langle F_R\rangle /\langle R\rangle )/4\pi$,
it seems advantageous to consider models with
larger $N$.  To see that this is not the case one can study the behavior
of $\beta_{N-1}$ and $\beta_N$ as functions of $N$ in the large-$N$
limit by explicitly minimizing the scalar potential.  A simple analysis
leads to the result that
\begin{equation}
	\beta_{N-1} \sim  - {(\beta_N)^{1/N}\over N + 1 +
		(N\beta_N)^{-2}} \quad{\rm with} \quad
	\beta_N \sim  {\cal O}(1/N).
\end{equation}
This implies in turn that $\beta_{N-1}\sim {\cal O}(1/N)$, as well as
\begin{equation}
\beta_{N-1} + (\beta_{N})^{1/N} /N \sim {\cal O}(1/N^2).
\label{eq:Nbehav}
\end{equation}
These equations show that both expressions
that enter the RG equations, $N^2(F_R/R)^2$ and $m_b^2$,
have the same $N$-scaling,
$$
m_b^2 \sim N^2 \left(\frac{F_R}{R}\right)^2 \sim {\cal O}(1/N^2),
$$
and thus no advantage is gained by going to the large-$N$ limit.

Therefore we find that all models in the $SU(N)\times SU(N-1)$ class,
suffer from the problem that the SSM squark and slepton $({\rm
mass})^2$ driven negative.\footnote{The situation improves if we embed
many ${\bf 5}$'s or larger representations into the global $SU(N-2)$
symmetry.  For instance, by embedding ${\bf 75}$ into the global
symmetry, which requires $N = 77$, we can enhance the gaugino
contribution by a factor of 50 relative to that of the $b$-scalars,
which might be barely enough.  However perturbativity of the SM gauge
couplings is then lost.}  Finally, note that the
$SU(N)\times SU(N-2)$ models of Poppitz and Trivedi \cite{PT2} also
suffer from the flaw that the two-loop contribution of light
$b_k$-like states drive SSM sfermion $({\rm mass})^2$
negative.\footnote{This has been independently realized by the
authors of Ref.~\cite{APT}.}

It is desired, therefore, to have a model of supersymmetry breaking
where none of the light degrees of freedom are charged under the (weakly
gauged) global symmetries, thus avoiding the dangers of this section
completely.  We explored all existing DSB
models available in literature and none of them appear to achieve the
three goals {\it while}\/ satisfying this requirement.  However, the
list of DSB models is growing rapidly recently, and non-renormalizable
models have not been explored extensively in literature.  We
believe that a continued effort along the scheme we propose will
result in a simple and phenomenologically viable model of gauge
mediation.

\section{Conclusion}
\setcounter{footnote}{0}
\setcounter{equation}{0}

We proposed a new scheme for the construction of simpler models of gauge
mediation.  The new scheme aims to simultaneously achieve the
following ambitious goals: (1) a much simpler structure by the direct
coupling of the standard model gauge groups to the DSB sector while
maintaining perturbative unification, (2) a natural suppression of
the supergravity contribution despite the high scale of supersymmetry
breaking, and (3) no fine-tuning of parameters or very
small coupling constant.  We found a modified class of DSB models based on
$SU(N)\times SU(N-1)$ which have classical flat
directions lifted quantum mechanically, and which allow the gauging of
an $SU(N-2)$ global symmetry.  Based on this new class of
models, we demonstrated that all the above goals can be achieved.

The basic idea for a successful direct
coupling is to employ models where at least one of the
classical flat directions $X$ is unlifted at the renormalizable level, but
is lifted after adding suitable non-renormalizable terms to the
superpotential.  The direction $X$ which is not lifted at the
renormalizable level then acquires a large
expectation value with small vacuum energy.  Therefore a natural
hierarchy $\langle F_{X} \rangle \gg \langle X \rangle^{2}$ is
achieved.  If the fields charged under the standard model gauge groups
acquire masses due to $\langle X \rangle$, their contribution to the
running of the SM gauge coupling constants appears only above $\langle
X\rangle$ and hence perturbative gauge unification is relatively
easy to preserve.  Even though the scales are much higher than in the
OGM models, the theories which we studied still naturally suppress the
supergravity contributions to the sfermion masses, and hence
squark degeneracy, the primary motivation of the gauge mediation
mechanism, is automatic.

Our requirements for a successful DSB model were given in
Section~\ref{sec:scheme} and are repeated here:
\begin{enumerate}
	\item It must accommodate a large global symmetry, such as $SU(5)$ or
	$SU(3)$.

	\item Some of the flat directions are unlifted at the
	renormalizable level.

	\item The addition of non-renormalizable operators lifts the flat
	directions and the model breaks supersymmetry.  The
	dimensionality of the non-renormalizable operators should
	satisfy the constraint $2 + \frac{14.4}{2+60/N} \lsim n \lsim
	9.2$.
\end{enumerate}

However, we found that there is a new type of problem which has not
been discussed before in literature. The particular models which we
employed generate supermultiplets below $10^{5}$~GeV charged under
the standard model gauge interactions, and their scalar components have
large soft-SUSY breaking masses of order $(10^{4}~\mbox{GeV})^{2}$.
They contribute to the renormalization group evolution of squark and
slepton masses at the two-loop level, and drive them negative at low
energies.  This problem unfortunately cannot be avoided by varying
the size of the gauge groups within the class of models we considered.

It is clear that the problem is rather specific to models which produce
light multiplets charged under the standard model gauge group.  It
is likely that there are models which do not produce such a
spectrum.  This consideration leads to a fourth requirement:
\begin{enumerate}
	\addtocounter{enumi}{3}
	\item Directions with non-trivial quantum numbers under
	the standard model gauge group are lifted at the renormalizable
	level to avoid light charged fields.
\end{enumerate}
While none of the existing DSB models available in literature appear to
satisfy the above four requirements, there has been great recent progress
in the construction of DSB models, and non-renormalizable models
arejust beginning to be extensively explored.  We strongly believe that a
continued effort along the the lines of our proposed
scheme will result in simple and phenomenologically viable
model of gauge mediation.
\section{Acknowledgments}
JMR wishes to thank K.~Babu and C.~Kolda for conversations and
S.~Trivedi for discussing with us the work
of \cite{APT} in which the logarithmically
enhanced negative contributions to the sfermion masses
are also considered.
\appendix
\section{Details of the $SU(N)\times SU(N-1)$ models}
\setcounter{footnote}{0}
\setcounter{equation}{0}

In this appendix we present some of the details of the analysis of
the model of section \ref{sec:model}.
We first show that the superpotential Eq.~(\ref{W})
lifts all classical flat directions except for $\phi \neq 0$
direction.  The complete set of gauge invariant operators is $Y_{i}^{I} =
L_{i}Q R^{I}$, $b_{I} = (RRRRRR)_{I}$, and ${\cal B}=\mbox{det}(LQ)$
with a constraint: $\epsilon_{i_{1}i_{2}\cdots
i_{6}}Y^{i_{1}I_{1}}Y^{i_{2}I_{2}}\cdots Y^{i_{6}I_{6}} \propto {\cal
B} \epsilon^{I_{1}I_{2}\cdots I_{6}I_{7}}b_{I_{7}}$.  We will show
that all operators vanish classically because of the conditions for a
supersymmetric vacuum.  The general
$SU(N)\times SU(N-1)$ case can be discussed in exact parallel.

The conditions for a supersymmetric vacuum are given by
\begin{eqnarray}
	\frac{\partial W}{\partial L_{i}} & = & \lambda Q R^{i} +
		\frac{g}{M} Q R^{6} \phi^{i} = 0,
	\label{FLi}  \\
	\frac{\partial W}{\partial L_{6}} & = & \lambda' Q R^{6} = 0,
	\label{FL6}  \\
	\frac{\partial W}{\partial \phi^{i}} & = &
		\frac{g}{M} L_{i} Q R^{6} + \frac{h}{M^{4}} b_{i}= 0,
	\label{Fphi} \\
	\frac{\partial W}{\partial R^{i}} & = & \lambda L_{i} Q
		+  \frac{\alpha}{M^{3}} (R^{5})_{6,i}
		+ \frac{h}{M^{4}} (R^{5})_{j,i} \phi^{j} = 0,
	\label{FRi}  \\
	\frac{\partial W}{\partial R^{6}} & = & \lambda' L_{6} Q
		+ \frac{g}{M} L_{i} Q \phi^{i}
		+ \frac{h}{M^{4}} (R^{5})_{j,6} \phi^{j} = 0,
	\label{FR6}  \\
	\frac{\partial W}{\partial R^{7}} & = &
		\frac{\alpha}{M^{3}} (R^{5})_{6,7}
		+ \frac{h}{M^{4}} (R^{5})_{j,7} \phi^{j} = 0,
	\label{FR7}  \\
	\frac{\partial W}{\partial Q} & = & \lambda L_{i} R^{i}
		+ \lambda' L_{6} R^{6}
		+ \frac{g}{M} L_{i} R^{6} \phi^{i} = 0.
	\label{FQ}
\end{eqnarray}
Eq.~(\ref{FL6})
requires $Q R^{6} = 0$, and therefore, $Y_{i}^{6}=0$ for
all $i$ Then Eq.~(\ref{FLi}) simplifies to $\lambda Q
R^{i} = 0$, and hence $Y_{j}^{i}=0$ for all $i,j=1, \cdots, 6$.
Once $Y_{j}^{6}$ are known to vanish, Eq.~(\ref{Fphi}) gives $b_{i} =
0$ for $i=1, \cdots, 5$.  Now we multiply Eq.~(\ref{FR7}) by
$R^{6}$.  The  second term vanishes, and we obtain $b_{7} =
0$.  By multiplying the same equation by $R^{7}$, we obtain $\alpha
b_{6} + h b_{j} \phi^{j}/M = 0$, but we know already $b_{j} = 0$
and conclude $b_{6} = 0$.  Therefore, all $b_{i}$, $b_{6}$, and
$b_{7}$ vanish.  Using Eq.~(\ref{FRi}) multiplied by $R^{7}$, the
second and third terms vanish, and we find
$Y_{i}^{7} = 0$.  Finally with Eq.~(\ref{FR6}) multiplied by $R^{7}$,
combined with vanishing of $Y_{i}^{7}$, we obtain
$Y_{6}^{7} = 0$.  Therefore, all $Y$'s vanish.  The last step is to
multiply Eq.~(\ref{FRi}) by $\phi^{i}$.  We find $\lambda L_{i} Q
\phi^{i} + \alpha (R^{5})_{6,i} \phi^{i}/M^{3} = 0$.  By using
Eq.~(\ref{FR6}), one can rewrite the condition as $\lambda L_{i} Q
\phi^{i} + (\alpha/h) (M \lambda' L_{6} Q + g L_{i} Q \phi^{i}) =
0$, and hence, $(\lambda + \alpha g/h) L_{i} Q \phi^{i} + (\alpha
\lambda'/h) M L_{6} Q = 0$.  If $\phi_{i} = 0$, $L_{6} Q = 0$ and
hence ${\cal B} = \mbox{det}(L_{i,6} Q) = 0$.  If $\phi_{i} \neq 0$, we
have a linear relation between $L_{i} Q$ and $L_{6} Q$ and hence
again ${\cal B} = 0$.  Therefore, ${\cal B}$ always vanishes.  This
concludes the proof that all gauge invariant polynomials vanish
classically except $\phi^{i}$.  As argued later, the $\phi$
flat direction gets lifted at the quantum level.

    The K\"{a}hler potential for $b_{I}$ at the classical level can be
obtained for any $N$ using the method of Poppitz and Randall \cite{PR},
where the heavy SU($N-1$) vector multiplet $V$ is integrated out classically
by setting it to its classical equation motion, leaving the
effective K\"{a}hler potential for the light moduli only. One
starts with the K\"{a}her potential for $R^{I}$ fields $K =
R^{\dagger} e^{V} R=\mbox{Tr} (e^{V} R R^{\dagger})$
, and requires the stationary condition of $K$
with respect to arbitrary variation of $V$(the $T^a$ are SU(N-1) generators)
:
\begin{equation}
0={d \over {d V^a}} \mbox{Tr} e^{V} R R^{\dagger}=\mbox{Tr}
T^a \int_0^1 dt e^{tV} R R^{\dagger} e^{(1-t)V}
\end{equation}
so
\begin{equation}
\int_0^1 dt e^{tV} R R^{\dagger} e^{(1-t)V} = c {\bf I}.
\end{equation}
It is easy to show that the above can only be satisfied if $U \equiv e^V$
commutes with $R R^{\dagger}$. Then, we must have $R R^{\dagger} U = c {\bf I}
\rightarrow \mbox{det}R R^{\dagger} = c^{N-1}$, since det$U=e^{\mbox{Tr}V}=1$.
Therefore, the K\"{a}her potential is
\begin{equation}
K=\mbox{Tr}R R^{\dagger} U = (N-1)c = (N-1) (\mbox{det} R R^{\dagger})^{{1 \over
{N-1}}}
\end{equation}
By writing down $\mbox{det} (R R^{\dagger})$ explicitly, we
obtain
\begin{eqnarray}
& &
	\mbox{det} (R R^{\dagger})
	= \frac{1}{(N-1)!}
	\epsilon^{\alpha_{1}\alpha_{2} \cdots \alpha_{(N-1)}}
	  \epsilon_{\beta_{1}\beta_{2} \cdots \beta_{(N-1)}}
	  (R^{I_{1}}_{\alpha_{1}} R^{*\beta_{1}}_{I_{1}})
	  (R^{I_{2}}_{\alpha_{2}} R^{*\beta_{2}}_{I_{2}})
	  \cdots
	  (R^{I_{(N-1)}}_{\alpha_{(N-1)}} R^{*\beta_{(N-1)}}_{I_{(N-1)}})
	   \nonumber \\
& &	= \frac{1}{(N-1)!}
	\epsilon^{I_{1}I_{2}\cdots I_{(N-1)}J} b_{J}
	  \epsilon_{I_{1}I_{2}\cdots I_{(N-1)}K} b^{*K}
	= b_{J} b^{*J}.
\end{eqnarray}
where we use the normalization $b_{J} = {1 \over {(N-1)!}}
\epsilon_{J I_{1} \cdots
I_{N-1}} R^{I_{1}} \cdots R^{I_{N-1}}$.  Therefore, we finally obtain
\begin{equation}
	K = (N-1) (b^{*J} b_{J})^{1/(N-1)}.
\end{equation}

In order to discuss the mass spectrum, we need to explicitly minimize
the potential and expand the theory around the vacuum.
Let us first determine the vacuum in the general
SU(N)$\times$SU(N$-$1) model \cite{PST1} without the $\phi$ field.  We
will show later that the classical flat direction $\phi\neq 0$ is
actually lifted and justify this treatment.  The superpotential is
\begin{equation}
	W = (\Lambda^{2N+1} (\mbox{det}' \lambda) b_{N})^{1/N} +
		\frac{\alpha}{M^{N-4}} b_{N-1} .
\end{equation}
We redefine $\Lambda$ to absorb the unimportant factor $\mbox{det}'
\lambda$ hereafter.  It is useful to rescale the fields as
\begin{equation}
	b_I = \alpha^{-N/(N-1)} \Lambda^{(2N+1)/(N-1)}
		M^{N(N-4)/(N-1)} \beta_I .
\end{equation}
By using the rescaled field $\beta_I$, the Lagrangian is given by
\begin{eqnarray}
	K &=& \left[ \alpha^{-N} \Lambda^{2N+1} M^{N(N-4)}
		\right]^{2/(N-1)^2}
	(N-1) (\beta^{*J} \beta_{J})^{1/(N-1)}, \\
	W &=& \left[\alpha^{-1} \Lambda^{2N+1} M^{N-4} \right]^{1/(N-1)}
		(\beta_{N}^{1/N} + \beta_{N-1}).
\end{eqnarray}
The matrix for the kinetic term and its inverse are given by
\begin{eqnarray}
	K^I_J &=& \left[ \alpha^{-N} \Lambda^{2N+1} M^{N(N-4)}
		\right]^{2/(N-1)^2}
	(\beta^\dagger \beta)^{-(N-2)/(N-1)}
	\left( \delta^i_j - \frac{N-2}{N-1}
		\frac{\beta \beta^\dagger}{\beta^\dagger\beta}
	\right) \\
	K^{-1I}_J &=&
		\left[ \alpha^{-N} \Lambda^{2N+1} M^{N(N-4)}
		\right]^{-2/(N-1)^2}
	(\beta^\dagger \beta)^{(N-2)/(N-1)}
	\left( \delta^i_j + (N-2)
		\frac{\beta \beta^\dagger}{\beta^\dagger\beta}
	\right)
\end{eqnarray}
Therefore, the potential is given by
\begin{equation}
	V = V_0 (\beta^\dagger \beta)^{(N-2)/(N-1)}
	\left( 1 + \frac{1}{N^2} (\beta^{*N} \beta_N)^{-(N-1)/N}
		+ \frac{N-2}{\beta^\dagger \beta}
		\left|\beta_{N-1} + \frac{1}{N} \beta_N^{1/N}\right|^2
	\right),
\end{equation}
where the overall scale of the potential is
\begin{equation}
	V_0 =
		\left[ \alpha^{1/(N-1)^2} \Lambda^2
			\left(\frac{\Lambda}{M}\right)^{(N-4)/(N-1)^2}
		\right]^2 .
\end{equation}
It is amusing to derive the above potential
starting with the original fields $R$; in the process,
we will directly determine $K^{-1}$. Restricting $R$ to lie on the
$D$ flat space, the potential is
\begin{equation}
V={{\partial W}\over {\partial R^{K}_{\alpha}}} {{\partial W^{*}}\over{\partial R^{* \alpha}_{K}}}
 ={{\partial W}\over {\partial b_{J}}} {{\partial W^{*}}\over {\partial b^{*I}}} K^{-1I}_{J}
\end{equation}
where
\begin{equation}
K^{-1I}_{J} = {{\partial b_J}\over{\partial R^{K}_{\alpha}}} {{\partial b^{*I}}\over{
\partial R^{* \alpha}_{K}}}
\end{equation}
A point on the $D$ flat space
is specified by values for $b_1,..,b_N$. We can always make a global rotation
to go to a basis where only $b_N$ is non-vanishing, and the D flat direction is $R^{i}_{i}=\rho$
for $i=1,..,N-1$ with all other $R$'s vanishing, and with $\rho^{(N-1)}=b_N$.
Now, if $R$ is taken to be an $N \times (N-1)$ matrix, $b_I$ is the determinant
of the matrix with the $I$'th column removed, and
${{\partial b_J}\over{R^{K}_{\alpha}}}$ is the determinant with the
$J,K$'th columns and the $\alpha$'th row removed. It is easy to compute
$K^{-1}$ in this basis, and we find
\begin{equation}
K^{-1I}_{J} = (\rho^* \rho)^{(N-2)}\left(\delta^I_J + (N-2) \delta^{I}_{N} \delta^{N}_{J}\right).
\end{equation}
But it is trivial to write this in a basis independent way:$(\rho^* \rho)=
(b^{\dagger} b)^{1\over(N-1)}$, $\delta^I_J$ is invariant and $\delta^{I}_{N} \delta^{N}_{J}$
is the projection operator onto the $b$ direction ${{b b^{\dagger}} \over {b^{\dagger} b}}$,so
\begin{equation}
K^{-1I}_{J}=(b^{\dagger} b)^{\frac {N-2}{N-1}}  \left( \delta^I_J + (N-2) {{b b^{\dagger}}\over{b^{\dagger} b}}
\right)
\end{equation}
exactly as before.

By expanding the above potential with respect to $\beta_{k}$ for
$k=1, \cdots, N-2$, and normalizing it correctly by the coefficient
of their kinetic term, the masses for these fields are given by
\begin{equation}
m_b^2 = m^2 (\beta^\dagger \beta)^{(N-3)/(N-1)}
\frac{N-2}{N-1} \left(
	1 + {1\over N^2} (\beta_N^\dagger \beta_N)^{-(N-1)/N}
	- \frac{1}{\beta^\dagger \beta} |\beta_{N-1}+\beta_N^{1/N}/N|^2
	\right)  ,
\end{equation}
where the overall scale is given by
\begin{equation}
	m \equiv \alpha^{(N+1)/(N-1)^2} \Lambda^{(2N+1)(N-3)/(N-1)^2}
            M^{-(N-4)(N+1)/(N-1)^2}  .
\end{equation}
Further coupling to the $\phi$ field, the potential reads
\begin{eqnarray}
\lefteqn{
	V = V_0 (\beta^\dagger \beta)^{(N-2)/(N-1)}
	\left( 1 + \frac{1}{N^2} (\beta^{*N} \beta_N)^{-(N-1)/N}
		+ \frac{h^2}{\alpha^{2}M^2} \phi^\dagger \phi
		+ \frac{N-2}{\beta^\dagger \beta} \right.} \nonumber \\
& &  \left.
		\left|\beta_{N-1} + \frac{1}{N} \beta_N^{1/N}
		+ \frac{h}{\alpha M} \phi^k \beta_k\right|^2
	\right)
	+ \left( \alpha^{-1} \Lambda^{2N+1} M^{N-4} \right)^{2/(N-1)}
		\frac{h^2}{\alpha^{2}M^2} (\beta^{*k} \beta_k),\nonumber \\
\end{eqnarray}
for $k=1,\cdots,N-2$.

Numerically minimizing the potential for the case $N=7$ gives the
location of the minimum
\begin{eqnarray}
	\beta_{6} &= & -0.0702 ,\\
	\beta_{7}   &=& 0.0791 .
\end{eqnarray}
%
By expanding the potential around the
minimum up to second order in $\beta_k$ and $\phi^k$, and writing
$\lambda \equiv h/\alpha M$ for simplicity, we obtain
\begin{eqnarray}
V &=& V_0 (0.070+ 4.409 \beta^{*k} \beta_k
		+ 0.309 \lambda (\beta_k \phi^k + \mbox{c.c.})
		+ 0.0236 \lambda^2 \phi_k^* \phi^k)  \nonumber \\
& &	+ \left( \alpha^{-1} \Lambda^{2N+1} M^{N-4} \right)^{2/(N-1)}
		\lambda^2 (\beta^{*k} \beta_k) .
\end{eqnarray}
For the purpose of proving that the $\beta_k$ and $\phi^k$ have positive
definite mass eigenvalues, we do not need to further rescale $\beta_k$
to make the K\"ahler potential canonical.  We can also drop the last
term since it is always an additional positive contribution to the
$\beta_k$ mass squared.  By taking the determinant of the mass matrix on
$(\beta_k, \phi^*_k)$ space, it is straightforward to prove that
$\phi^k=\beta_k=0$ is a stable minimum for any values of $\lambda$ and
$\alpha$. Thus, the model keeps an SU(5) symmetry intact at the
minimum of the potential. Since only $b_6,b_7$ are non-zero at the minimum,
the corresponding $R$ on $D$ flat space has the form $R^{i}_{i}=R$ for
$i=1,..,5$,$R^{6}_{6}=\psi,R^{7}_{6}=\chi$ with $|\psi|^2 + |\chi|^2 = |R|^2$,
and all other $R$'s vanishing. The unbroken SU(5) is the diagonal subgroup of the
original global SU(5) and the gauged SU(6) symmetries. The $Q,L$ fields then
contain 7 pairs of $({\bf 5} + {\bf 5^*})$ under SU(5), with supersymmetric
mass $R$ and supersymmetry breaking bilinear $F_R \equiv F_{R^i_i}$, and
they can mediate supersymmetry breaking to the ordinary sector.

We now determine $R$ and $F_R$ in terms of
$b_{N,N-1}$ for general N:
\begin{equation}
R=(b_{N-1}^2 + b_{N}^2)^{1/2(N-1)},
\end{equation}
\begin{equation}
F_R = {{\partial W}\over {\partial R^i_i}} = \Lambda^{(2N + 1)/N} {1\over N} b_N^{-(N-1)/N}
\frac{\partial b_N}{\partial R^i_i} + \frac{\alpha}{M^3} \frac{\partial b_{N-1}}{\partial R^i_i}.
\end{equation}
But $\frac{\partial b_N}{\partial R^i_i}=b_N/R,\frac{\partial b_{N-1}}{\partial R^i_i}=b_{N-1}/R$,
so
\begin{equation}
F_R = 1/R \left(\Lambda^{(2N + 1)/N} {1\over N} b_N^{1/N} +
\frac{\alpha}{M^3} b_{N-1} \right),
\end{equation}
and we find easily
\begin{equation}
\frac {F_R}{R} = m \frac{(\beta_{N-1} + \beta_N^{1/N}/N)}{(\beta^{\dagger} \beta)^{1/(N-1)}}
\end{equation}
with $m$ as given in (A.22).

Finally, we demonstrate that the classical flat direction
$\phi\neq 0$ is lifted quantum mechanically.  As clear from
the analysis above, the vacuum energy increase as
$\alpha^{2N/(N-1)^2}$.  In the presence of $\phi \neq 0$, one can
perform a rotation in the flavor space so that the $N-1$-dimensional
vector $(\lambda \phi^{I}, \alpha)$ has a value only in the $N-1$-th
component.  Note that such a rotation keeps the non-perturbative
superpotential $\propto b_{N}^{1/N}$ intact.  It effectively increases
the value of $\alpha$ to $\alpha_{\rm new} =
\sqrt{|\alpha|^2 + |\lambda|^{2} \phi^\dagger \phi} \geq
|\alpha|$.  Therefore, the classical flat direction $\phi$ is
lifted quantum mechanically and it develops a stable minimum when
$\phi$ is driven back to the origin.

\end{document}